# *In-situ* electrochemical fabrication of natural contacts on single nanowires

*Wenhao Wu, J. B. DiMaria,*[a] *and Han G. Yoo*

*Department of Physics and Astronomy, University of Rochester, Rochester, NY 14627*

*Shanlin Pan and L. J. Rothberg*

*Department of Chemistry, University of Rochester, Rochester, NY 14627*

*Yong Zhang*

*Department of Chemical Engineering, University of Rochester, Rochester, NY 14627*

**Abstract**

We report a template-based *in-situ* electrochemical method for fabricating *natural* electric contacts on single nanowires using a pair of cross-patterned electrodes. Such electric contacts are highly stable upon thermal cycling between room temperature and milli-Kelvin temperatures. Direct imaging of the single-nanowire contacts using scanning electron microscopy is also demonstrated.

[a] Permanent Address: Physics Department, University of Connecticut, Storrs, CT 06269

Electrochemical deposition of nanowires based on a porous membrane template has been adopted by many researchers for the fabrication of a wide variety of nanowire materials with potential applications in a broad range of areas.[1,2] Electrochemical deposition is accomplished by coating one face of the membrane with a thick (~ 200 nm) metal (Au or Ag) film and using this film as a cathode (the back electrode) for electroplating to form an organized array of nanowires in the porous membrane.

For various nanowires fabricated using this method, transport studies have mainly been carried out on large arrays of nanowires using two electrodes fabricated on the opposite faces of the membranes. This method does not allow measurements of the absolute value of the resistance of a single wire. In addition, it has been difficult to make reliable electric contact on the nanowires with a low contact resistance, possibly due to a poor surface layer or oxidation of the nanowires, as demonstrated by a recent study.[3] Our measurements on arrays of nanowires of Au, Ag, Bi, Co, Cu, Fe, and Pb deposited in porous alumina membranes (pore diameter 200-250 nm and length ~ 60 μm) have also yielded inconsistent results. Using electrodes of 0.5 mm$^2$ in area, we typically measure resistance values that are orders of magnitude larger than what we would expect based on the bulk resistivity values.

Recently, a number of methods[3-11] has been developed to fabricate electric contacts on single nanowires. One of them is an *in-situ* electrochemical technique[6-11] that has allowed fabrication of contacts on single nanowires based on a rather simple procedure. This method uses a relatively thin (~ 50 nm) Au film evaporated on the membrane face exposed to the electrolyte to form a front electrode. This front electrode covers a portion of the pores to reduce the density of the pores. During electroplating, a sharp increase in the current can be detected as the first nanowire emerges from a pore and makes a contact with the front electrode. A feedback mechanism immediately terminates electroplating to avoid the growth of other nanowires. This method thus can fabricate contacts to only one or a few nanowires, and it works easily for membranes with a low density of pores. However, it has not been possible to directly investigate the nature of such single-nanowire contacts by any microscopy. The stability of the contact is also uncertain for some samples, due to the immediate interruption of electroplating following a sharp increase in the current.

Here we report a new *in-situ* electrochemical method for fabricating highly reliable *natural* electric contacts on single nanowires. We purchase porous membranes from Whatman International, Ltd. The typical pore diameter is 200-250 nm. The length of the pores is 60 μm. This method is based on a pair of crossed narrow (~ 1.0 mm) metal (Au or Ag) film electrodes evaporated on the opposite faces of a membrane, as shown in Fig. 1. Both electrodes are thick enough (~ 200-300 nm) to completely block all the pores covered by them. The back electrode serves as the cathode in electroplating. The anode is located a few centimeters away from the other face of the membrane. This electrode pattern greatly reduces the number of pores relevant to making contacts to the front electrode, since only pores near the two edges along the front electrode at the crossing of the two electrodes are relevant. During electroplating, the current through the electrolyte cell first increases slowly as the pores are getting filled. When the fastest growing nanowire emerges from a pore, a significant increase in current is observed, as indicated by the left arrow in the upper frame of Fig 2. This is due to the quick formation of a large mushroom head on the tip of the emerging nanowire as its growth is no longer limited by diffusion of ions into the pore. The mushroom head closest to the front electrode eventually contact the edge of the front electrode, resulting in a sharp increase in the current (I) and a sharp drop in the potential difference (V) between the front and the back electrodes. If electroplating is not terminated, a series of current and potential steps can result, indicating a number of mushroom heads contacting the front electrode. In the upper frame of Fig. 2, we indicate two such steps by the second and the third arrows along the measured voltage and current traces. In the lower frame of Fig. 2, we show a scanning electron microscopy (SEM) picture taken on the same sample on which the traces in the upper frame of Fig. 2 were measured. In this picture, there are a few mushroom heads observable, with two of them (in white boxes) clearly in contact with the front electrode. These two contacts correspond well with the two sharp steps observed in the current and voltage traces in the upper frame in Fig. 2. Therefore, a reliable electric contact between a single mushroom head and the front electrode can be fabricated if electroplating is terminated before the second step in current and voltage is observed. We note that, by applying an appropriate potential difference ($V_0 = 2.5 - 3.0$ V for Co and Ag deposition) across the anode and the cathode, electroplating can be initiated simultaneously inside the pores and on the surface of the front electrode. This is confirmed by studies of the chemical composition of the mushroom heads and the front electrode using energy dispersive analysis of X-rays (EDAX). As a result, the mushroom heads and the front electrode form *natural* contacts between identically electroplated materials, with a zero contact resistance.

We believe this method, in fact, forms a contact between a single nanowire and the front electrode. Based on the low density of mushroom heads seen in SEM pictures, it is unlikely that a typical mushroom head can be in contact with more than one nanowire. We have also carried out SEM investigations of the nanowires under the mushroom heads after the alumina membrane has been partially etched away using a 1M NaOH solution for 5-10 min. In the upper frame of Fig. 3, we show one SEM picture of a Co mushroom head with three partially exposed Co nanowires below it. It appears that only one of the three nanowires is in contact with the mushroom head. In the lower frame of Fig. 3, we show another SEM picture of an Ag mushroom head on clearly a single Ag nanowire.

In Fig. 4, we plot resistance data for single-contact Co and Ag nanowire samples measured in a quasi-four-probe configuration using the crossed back and front electrodes. Such contacts are very stable upon thermal cycling between room temperature and milli-Kelvin temperatures. The pore diameter was about 200-250 nm, and the length of the pores was 60 μm. Using the bulk resistivity values of Co and Ag near room temperature, $5.8 \times 10^{-6}$ Ωcm and $1.59 \times 10^{-6}$ Ωcm, respectively, we estimate at room temperature resistance values of 70-110 Ω for a single Co nanowire and 19-30 Ω for a single Ag nanowire. The data in Fig. 4 show relatively small variations in the resistance values measured on a few single-contact samples, within the range of the estimates. This further supports our conclusion that the contact is indeed made on a single nanowire. Compared with results in an earlier report,[11] the ration R(300K)/R(10K) for our Co samples is twice as large, while the overall temperature dependence of the resistance is very similar. Above 100 K, the resistance is nearly linear in T, with deviation from a linear dependence observable above 230 K. This behavior is expected for bulk Co,[12] with a linear contribution due to phonon scattering and an additional spin-disorder resistivity at high temperatures. The resistance of our Ag samples demonstrates a linear T dependence over a very broad temperature range. The Debye temperature for Ag is 215 K. A linear T dependence is expected near and above the Debye temperature.

In conclusion, we have developed an *in-situ* electrochemical method for the fabrication of *natural* electric contacts on single nanowires, using a pair of crossed electrodes. We have imaged the single-nanowire contacts using SEM. Such contacts are very stable upon thermal cycling down to low temperatures. We have obtained consistent resistance data from single Co and Ag nanowires. This method can be applied to fabricate single-nanowire contacts for a variety of materials.

This material is based upon work supported in part by NSF Grant Nos. DMR-0305428 and PHY-0242483. L. R. and Y. Z. were supported by NSF Grant No. DMR-030944. S. P. was supported by Infotonics Grant No. DE-FG02-02ER63410.A000.

**Figure Captions**

Figure 1  Sketch illustrating *in-situ* fabrication of natural contacts on single nanowires. The back electrode (vertically oriented) is the cathode. The current flowing through the cathode and the anode is monitored with a current meter. A voltmeter is used to measure the potential difference between the cathode and the front electrode (horizontally oriented). During electroplating, nanowires grow inside the pores starting from the cathode. Fast growing nanowires emerge from the pores to quickly form large mushroom heads. The mushroom head closest to the front electrode will contact the front electrode first.

Figure 2  Upper frame shows the voltage difference between the front electrode and the cathode, as well as the current through the electrolyte cell, versus time. The first arrow on the left indicates a significant decrease in the voltage and increase in the current, corresponding to the emergence of nanowires from the pores. The other two arrows indicate two lager steps in voltage and current, corresponding to contacts made between the front electrode and two mushroom heads, before deposition is terminated. The lower frame is one SEM picture taken on the same sample. A number of mushroom heads are observable in this picture, with two of them (in white boxes) clearly in contact with the front electrode (the flat region on the right). The bright regions are due to charging of the membrane.

Figure 3    SEM pictures of two mushroom heads taken after the porous membrane templates have been partially etched away using a NaOH solution. Picture on top was taken from a Co sample. The white, cloud-like material, formed after the sample was pulled out of the NaOH solution and dried, is $Al_2O_3$, based on EDAX analysis. The bottom picture shows an Ag mushroom head on a (bent) single Ag nanowire.

Figure 4    Resistance data for four single-contact Co nanowire samples (part A) and two Ag nanowire samples (part B) versus the temperature. Four of the samples were fabricated within a period of a week, while the other two samples represented by solid triangles were fabricated one month earlier. The straight lines are guides to eye.

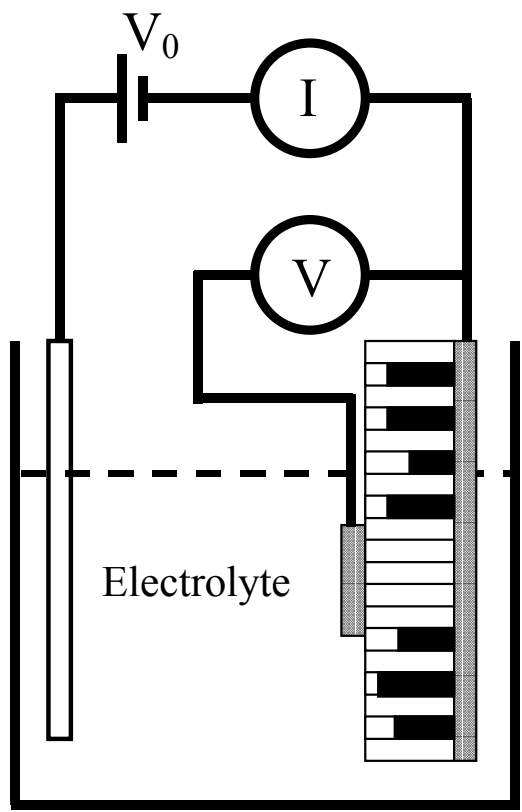

Figure 1        Wenhao Wu *et al*.

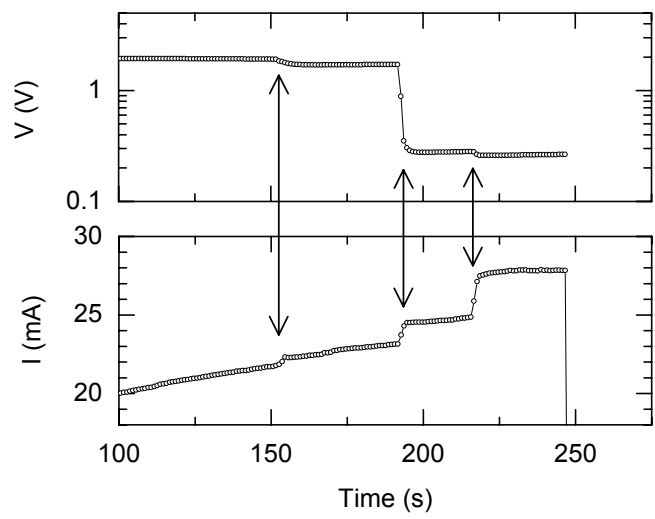

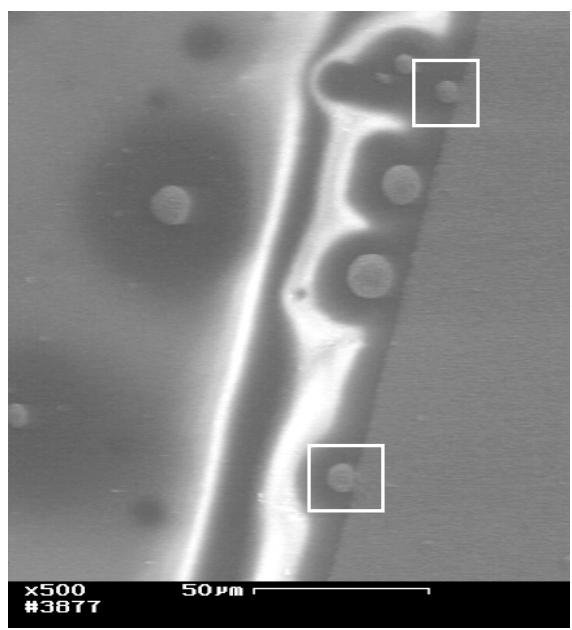

Figure 2 Wenhao Wu *et al.*

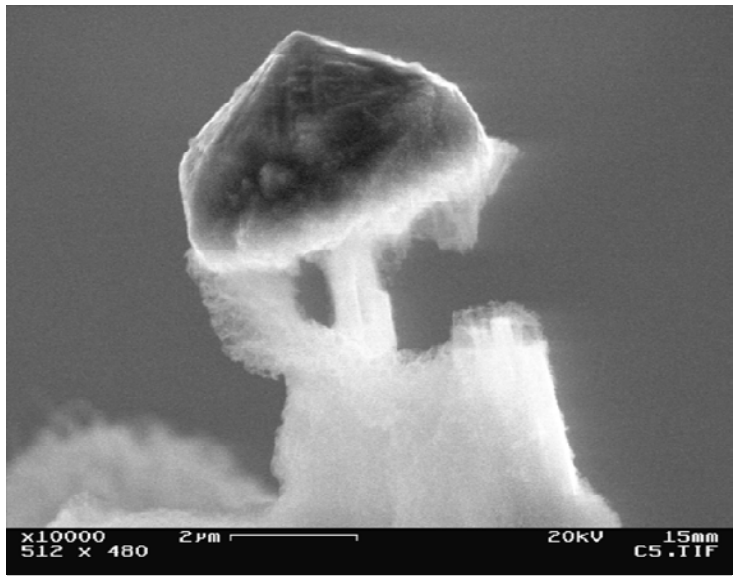

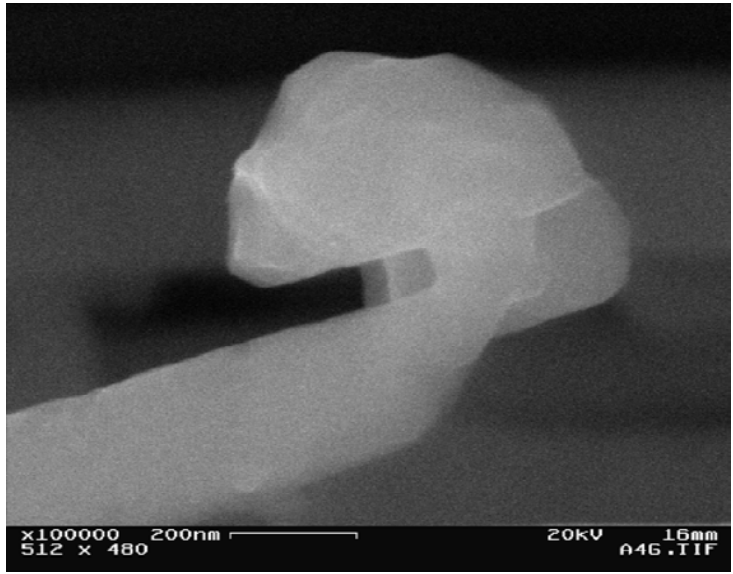

Figure 3  Wenhao Wu *et al.*

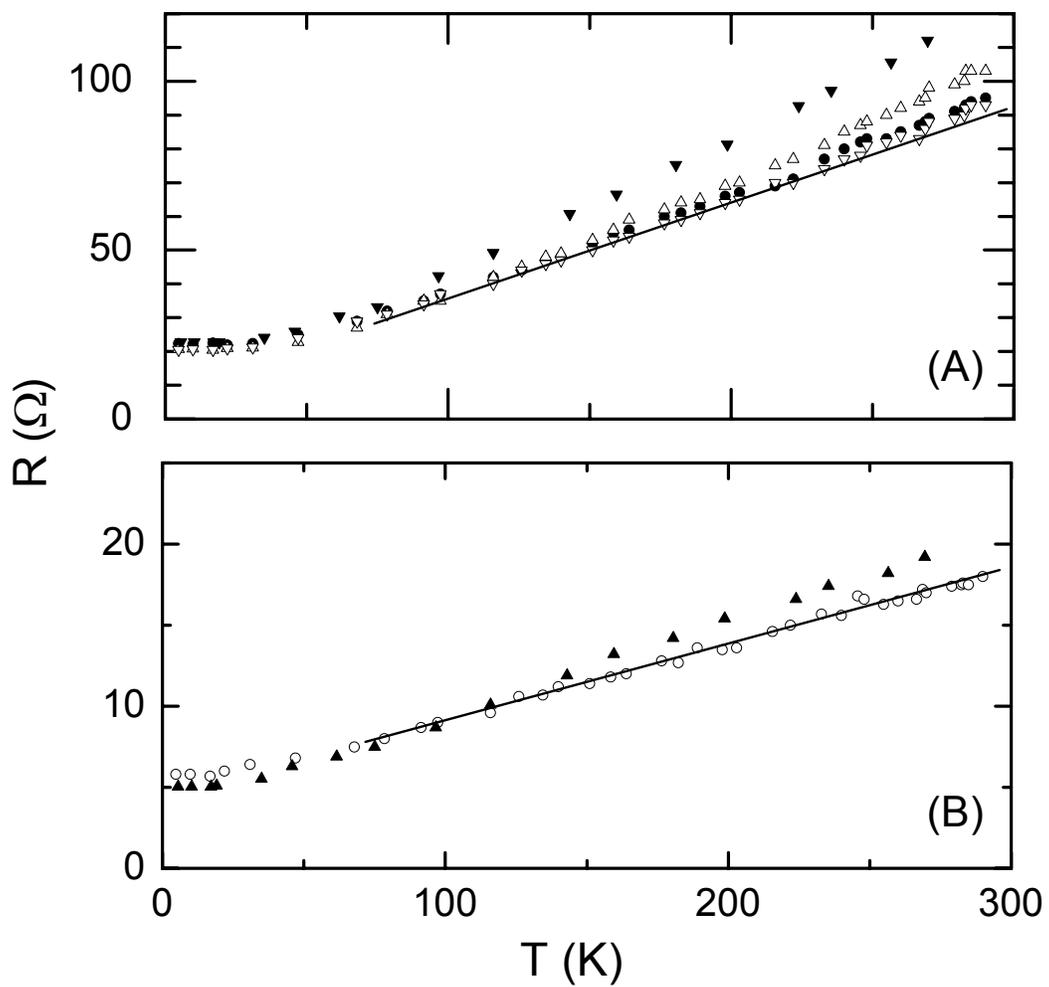

Figure 4    Wenhao Wu *et al.*

13